\documentstyle[prl,aps]{revtex}

\begin{document}

\title{Nucleon-Quarkonium Elastic Scattering And The Gluon Contribution To Nucleon
Spin. }
\author{{\bf Pervez Hoodbhoy} \\
Department of Physics, Quaid-e-Azam University, \\
Islamabad 45320, Pakistan.}

\maketitle

\begin{abstract}
It is shown that the amplitude for the scattering of a heavy quarkonium
system from a nucleon near threshold is completely determined by the
fraction of angular momentum, as well as linear momentum, carried by gluons
in the nucleon. A form for the quarkonium-nucleon non-relativistic potential
is derived.
\end{abstract}

Heavy quark-antiquark systems are essentially hydrogenic, have a size that
varies in inverse proportion to the heavy quark mass, and therefore have an
interaction with light hadrons that can be systematically and rigourously
analyzed using QCD and the operator product expansion \cite
{voloshin,gottfried}. The possibility that the attractive force between the $%
c\overline{c}$ system and nucleons may be strong enough to create a bound
state has been speculated upon for some time now\cite{brodsky,luke}. More
recently, Brodsky and Miller \cite{miller} have confirmed that those various
contaminations to the nucleon-$J/\psi $ force that are not computable from
first-principles are numerically small. They have also suggested using the
exclusive reaction $\pi ^{+}+D\rightarrow J/\psi +p+p$ as a means to
investigate low-energy nucleon-$J/\psi $ elastic scattering.

In this note we suggest that slightly off-forward \smallskip nucleon-$J/\psi 
$ elastic scattering can be used to probe the gluon angular momentum content
of the nucleon. This follows the proposal by Ji \cite{ji} who suggested
using deeply virtual compton scattering as a means of measuring the quark
orbital contribution to the nucleon's spin.

We begin with the covariantized version of the expression derived by Bhanot
and Peskin \cite{bhanot} for the scattering amplitude, 
\begin{equation}
{\cal M}=\frac{1}{2}\sum_{n=2}^{\infty }d_{n}a_{0}^{3}\varepsilon
_{0}^{2-n}v_{\mu _{1}}\cdots v_{\mu _{n}}\langle p^{\prime }|O^{\mu
_{1}\cdots \mu _{n}}|p\rangle  \label{one}
\end{equation}
In the above, $n$ runs over the even integers, $\varepsilon _{0}$ is the
binding energy of the quarkonium, $a_{0}$ is its the Bohr radius, and $d_{n}$
are exactly calculable coefficients \cite{bhanot} in the large $m_{Q}$
limit. The operator $O^{\mu _{1}\cdots \mu _{n}}$ is, 
\begin{equation}
O^{\mu _{1}\cdots \mu _{n}}=F^{\mu _{1}\rho }iD^{\mu _{2}}\cdots iD^{\mu
_{n}}F_{\rho }^{\vspace{-0.04in}\vspace{-0.25in}\;\mu _{n}}.
\end{equation}
>From this we are led to define, 
\begin{equation}
T^{\mu _{1}\cdots \mu _{n}}=F^{(\mu _{1}\rho }\stackrel{\longleftrightarrow 
}{iD^{\mu _{2}}}\cdots \stackrel{\longleftrightarrow }{iD^{\mu _{n}}}F_{\rho
}^{\;\vspace{-0.04in}\vspace{-0.25in}\mu _{n})}.
\end{equation}
The brackets refer to symmetrization in all n indices and subtraction of
traces. This procedure makes the above operator, which could be called the
generalized gluon stress-energy tensor, transform according to a definite
representation of the Lorentz group and have twist equal to two. The matrix
elements of $T^{\mu _{1}\cdots \mu _{n}}$ and $O^{\mu _{1}\cdots \mu _{n}}$
between hadron states are equal up to terms of $O(m/E)$ where $E$ is the
energy of the proton in the quarkonium rest frame. For infinitely massive
quarks, the 4-velocity of the quarkonium $v^{\mu }$ is conserved in the
scattering although its momentum is not.

The matrix element of the traceless tensor $T^{\mu _{1}\cdots \mu _{n}}$
taken between states of unequal momentum can be written in complete
generality as\cite{hoodbhoy},

\begin{eqnarray}
\langle p^{\prime }|T^{\mu _{1}\cdots \mu _{n}}|p\rangle  &=&\;\overline{u}%
(p^{\prime }s^{\prime })\gamma ^{(\mu _{1}}u(ps)\sum_{i=0}^{[\frac{n-1}{2}%
]}A_{n,2i}(t)\Delta ^{\mu _{2}}\cdots \Delta ^{\mu _{2i+1}}P^{\mu
_{2i+2}}\cdots P^{\mu _{n})}  \nonumber \\
&&+\overline{u}(p^{\prime }s^{\prime })\frac{\sigma ^{(\mu _{1}\alpha
}i\Delta _{\alpha }}{2m}u(ps)\sum_{i=0}^{[\frac{n-1}{2}]}B_{n,2i}(t)\Delta
^{\mu _{2}}\cdots \Delta ^{\mu _{2i+1}}P^{\mu _{2i+2}}\cdots P^{\mu _{n})}+ 
\nonumber \\
&&+\overline{u}(p^{\prime }s^{\prime })u(ps)\frac{1}{m}C_{n}(t)\Delta ^{(\mu
_{1}}\cdots \Delta ^{\mu _{n})}.  \label{N1}
\end{eqnarray}
The last term exists for even $n$ only. In the above, 
\begin{eqnarray}
P^{\mu } &=&\frac{1}{2}(p^{\prime }+p)^{\mu } \\
\Delta ^{\mu } &=&(p^{\prime }-p)^{\mu } \\
t &=&\Delta ^{2}.
\end{eqnarray}
While our discussion has been couched in a covariant language, the greatest
insight is obtained by working in the quarkonium rest frame. Considering the
quarkonium to be infinitely massive, the nucleon scatters elastically ($%
E^{\prime }=E=\sqrt{\vec{p}^{\;2}+m^{2}})$ through an angle $\theta $ in
terms of which, 
\begin{equation}
t=-4\vec{p}^{\;2}\sin ^{2}(\theta /2).
\end{equation}
We shall only be interested in the special kinematic region where $\theta $
is sufficiently small so that $t\ll m^{2}.$ Under the assumption that the
matrix elements of $T^{\mu _{1}\cdots \mu _{n}}$ and $O^{\mu _{1}\cdots \mu
_{n}}$ are approximately equal, which will be true if $\gamma =E/m\gg 1$, a
simple calculation yields,

\begin{equation}
{\cal M}=\frac{1}{2}a_{0}^{3}\varepsilon _{0}^{2}\sum_{n=2}^{\infty
}d_{n}\left( \frac{\gamma m}{\varepsilon _{0}}\right) ^{n}\left\{
2A_{n0}+\gamma B_{n0}\langle i\sigma \cdot \hat{n}\rangle \theta \right\} ,
\label{M1}
\end{equation}
where, 
\begin{equation}
\hat{n}=\frac{\vec{p}^{^{\;\prime }}\wedge \vec{p}}{\parallel \vec{p}%
^{\;^{\prime }}\wedge \vec{p}\parallel }
\end{equation}
is the vector normal to the scattering plane and $A_{n0}\equiv A_{n0}(t=0),$ 
$B_{n0}\equiv B_{n0}(t=0).$ The quantity $\langle i\sigma \cdot \hat{n}%
\rangle $ is $\pm 1$ if the initial and final nucleon spins are flipped and
zero otherwise, while the first term (proportional to $A_{n0})$ vanishes for
a spin-flip.

In principle, Eq. (\ref{M1}) allows for the extraction of $A_{n0}$ and $%
B_{n0}$, which are quantities of significant physical importance. From Eq. (%
\ref{N1}) it follows that $A_{n0}$ is the $n$'th moment of the usual
(forward) gluon distribution $G(x),$

\begin{equation}
A_{n0}=\int\limits_{-1}^{+1}dxx^{n-1}G(x).
\end{equation}
It is evident that $A_{20}$ has special significance as the momentum
fraction carried by gluons in a proton ($A_{20}${\boldmath$\approx$} $0.5$). 
To see
the significance of $B_{n0},$ define a gluon angular momentum from the
following generalization of the gluon angular momentum tensor,

\begin{equation}
{\bf M}^{\alpha \beta \mu _{1}\cdots \mu _{n}}(x)=x^{\alpha }O^{\beta \mu
_{1}\cdots \mu _{n}}-x^{\beta }O^{\alpha \mu _{1}\cdots \mu _{n}}-traces.
\label{Ang}
\end{equation}
There is only one ``reduced matrix element'' of the above, denoted by $J_{n}$
below\cite{hoodbhoy},

\begin{equation}
\langle PS|\int d^{4}x{\bf M}^{\alpha \beta \mu _{1}\cdots \mu
_{n}}(x)|PS\rangle =2J_{n}\frac{2S_{\rho }P_{\sigma }}{(n+1)m^{2}}\left\{
\varepsilon ^{\alpha \beta \rho \sigma }P^{\mu _{1}}\cdots P^{\mu
_{n}}+\cdots \right\} ,  \label{Jn}
\end{equation}
where the ellipses denote antisymmetrization with respect to the indices $%
\alpha ,\beta $ and subtraction of traces. One can therefore {\it define} a
gluon angular momentum distribution $J(x)$ through, 
\begin{equation}
J_{n}=\int\limits_{-1}^{+1}dxx^{n-1}J(x).
\end{equation}
Inserting Eq. (\ref{Ang}) into Eq. (\ref{Jn}) after taking the forward limit
appropriately yields, 
\begin{equation}
J_{n}=\frac{1}{2}\left( A_{n0}+B_{n0}\right) .
\end{equation}
In particular, for $n=2$, the sum $A_{20}+B_{20}$ is twice the angular
momentum carried by gluons in the nucleon.

Unfortunately, the kinematic conditions required for extracting all the $%
B_{n0}$ from Eq. (\ref{M1}) are unachievable --- it is difficult to see how
one can keep $t\ll m^{2}$ while maintaining $\gamma \gg 1$, which is
necessary if the matrix elements of difference $T^{\mu _{1}\cdots \mu
_{n}}-O^{\mu _{1}\cdots \mu _{n}}$ are to be ignorable. However, for $n=2$,
it is possible to use the trick of the QCD scale anomaly \cite{kaidalov,luke}%
to calculate the trace part. Essentially this consists of writing the matrix
element of the dominant interaction operator at the threshold energy using,

\begin{equation}
\langle p^{\prime }|E\cdot E|p\rangle =v_{\mu _{1}}v_{\mu _{2}}\langle
p^{\prime }|T^{\mu _{1}\mu _{2}}|p\rangle -\frac{g^{3}}{2\beta (g)}\langle
p^{\prime }|\theta _{\mu }^{\mu }|p\rangle ,
\end{equation}
All gluon colours are summed over in the product of the two colour electric
fields, and the scale anomaly has been used to relate the trace of the
stress energy tensor $\theta ^{\mu \nu }$ with the square of the gluon field
strength and the QCD beta function $\beta (g)$, 
\begin{eqnarray}
\partial _{\mu }j_{D}^{\mu } &=&\theta _{\mu }^{\mu }=\frac{\beta (g)}{2g^{3}%
}F^{\mu \nu }F_{\mu \nu }, \\
\beta (g) &=&-\frac{9g^{3}}{16\pi ^{2}}.
\end{eqnarray}
A straightforward calculation (keeping only the $n=2$ term in Eq.1) yields
the following simple expression for the small angle scattering amplitude, 
\begin{eqnarray}
{\cal M} &=&\frac{1}{6}d_{2}m^{2}a_{0}^{3}\left[ \frac{16\pi ^{2}}{9}%
+2A_{20}(\gamma ^{2}-\frac{1}{4})\right] +  \nonumber \\
&&\frac{1}{6}d_{2}m^{2}a_{0}^{3}(\gamma -1)\left[ (\gamma +\frac{1}{4}%
)A_{20}+\gamma (\gamma +1)B_{20}\right] \langle i\sigma \cdot \hat{n}\rangle
\theta .
\end{eqnarray}
In the above $d_{2}=28\pi /27.$ This formula is valid provided the velocity
of the nucleon in the quarkonium rest frame is small (so that all but the
first term in Eq. \ref{one} can be discarded), and the scattering angle $%
\theta $ is also sufficiently small. The normalization of the first term
agrees with that of Kaidalov and Volkovitsky\cite{kaidalov} but disagrees
with that of Luke, Manohar, and Savage \cite{luke}. Brodsky and Miller \cite
{miller} essentially propose to test the forward $(\theta =0)$ part of the
above equation using the exclusive reaction $\pi ^{+}+D\rightarrow J/\psi
+p+p.$

One can ask what non-relativistic nucleon-quarkonium potential $V_{\Phi p}(%
\vec{r})$ will reproduce the scattering amplitude ${\cal M}$ close to
threshold ($\gamma \rightarrow 1$) when inserted into the Schrodinger
equation and used in the Born approximation. It is easily seen that $V_{\Phi
p}(\vec{r})$ below fulfils this, 
\begin{equation}
V_{\Phi p}(\vec{r})=V_{1}(r)+\vec{\nabla}V_{2}(r)\cdot (i\vec{\sigma}\wedge 
\vec{\nabla}),
\end{equation}
where $V_{1}(r)$ and $V_{2}(r)$ are constrained to obey, 
\begin{eqnarray}
\int\limits_{0}^{\infty }drr^{2}V_{1}(r) &=&\frac{d_{2}ma_{0}^{3}}{48\pi }%
\left( \frac{16\pi ^{2}}{9}+\frac{3}{2}A_{20}\right) \\
\int\limits_{0}^{\infty }drr^{2}V_{2}(r) &=&\frac{d_{2}a_{0}^{3}}{48\pi m}%
\left( \frac{5}{8}A_{20}+B_{20}\right) .
\end{eqnarray}
We stress that no model parameters are involved; the above is a rigourous
result of QCD in the $m_{Q}\rightarrow \infty $ limit.

To conclude, it was shown that the amplitude for the scattering of a heavy
quarkonium system from a nucleon at threshold is completely determined by
the fraction of angular momentum, as well as linear momentum, carried by
gluons in the nucleon. Totally exclusive experiments involving the elastic
low energy scattering of quarkonium from nuclei may therefore be a means of
investigating the gluonic angular momentum component of nucleons. Although
going from nucleons to nuclei necessarily brings in the nuclear
wavefunction, for light enough nuclei like the deuteron this wavefunction is
probably sufficiently well known to embark on a meaningful experiment.

\medskip

{\Large Acknowledgements}

The author thanks Stanley Brodsky and Xiangdong Ji for encouragement and
comments. This work was supported by the Pakistan Science Foundation.

\end{document}